\documentclass[10pt,final,letterpaper,twocolumn,romanappendices]{IEEEtran}
\usepackage[dvips]{graphicx}
\usepackage{cite}
\usepackage{epsfig}
\usepackage{amsmath,amssymb,amsfonts,amstext,amsbsy,amsopn,eucal,dsfont}
\usepackage{psfrag}
\usepackage{subfigure}
\usepackage{array}
\usepackage{color}

\newtheorem{theorem}{\textbf{Theorem}}

\newtheorem{lemma}{\textbf{Lemma}}

\newcommand{\defn}{\triangleq}
\newcommand{\dif}{\textmd{d}}

\begin{document}

\title{\huge Coexisting Success Probability and Throughput of Multi-RAT Wireless Networks with Unlicensed Band Access}

\author{Xu Ding, Chun-Hung Liu, Li-Chun Wang and Xiaohui Zhao
\thanks{Xu Ding and Xiaohui Zhao are with the Department of Communication Engineering, Jilin University, Changchun China. C.-H. Liu and L.-C. Wang are with the Department of Electrical and Computer Engineering, National Chiao Tung University, Hsinchu Taiwan. The contact author is Dr. Liu  (Email: chungliu@nctu.edu.tw). Manuscript date: \today.}}

\maketitle

\begin{abstract}
In this letter, the coexisting success probability and throughput of a wireless network consisting of multiple subnetworks of different radio access technologies (RATs) is investigated. The coexisting success probability that is defined as the average of all success probabilities of all subnetworks is found in closed-form and it will be shown to have the concavity over the number of channels in the unlicensed band. The optimal deployment densities of all different RATs access points (APs) that maximize the coexisting success probability are shown to exist and can be found under the derived constraint on network parameters. The coexisting throughput is defined as the per-channel sum of all spectrum efficiencies of all subnetworks and numerical results show that it is significantly higher than the throughput of the unlicensed band only accessed by WiFi APs.
\end{abstract}

\begin{IEEEkeywords}
Coexistence, success probability, unlicensed band, Poisson point process, Mart\'{e}n hard-core point process
\end{IEEEkeywords}

\section{Introduction}
With the proliferation of wireless smart handsets, cellular data traffic is expected to considerably grow  to meet huge and different throughput demands entailed by versatile networking services. To make cellular networks jump over the high throughput hurdle due to limited licensed spectrum (band), deploying small cells with unlicensed band access is a promising mean of mitigating the spectrum crunch crisis. However, when small cells extend their services to the unlicensed band, how to mitigate or even eliminate interfering between small cells and the existing unlicensed access points (APs) (such as WiFi APs) is extremely pivotal since it is the root of reaching the well-coexisting goal of a wireless network with multiple radio access technologies (RATs) \cite{HZXCWGSW15}.  

The most important and fundamental issue that intrinsically exists in a wireless network consisting of multiple subnetworks of different RATs is the coexistence performance between these subnetworks.  Specifically, the success probability (i.e., the counterpart of outage probability) and throughput of all the coexisting subnetworks are the two most important performance metrics that need to be first investigated since they can indicate whether or not coexistence essentially benefits the entire network. To simply and thoroughly delve the success probability and throughput performances in the multi-RAT network, the concept of  coexisting success probability is first proposed in this letter for the small cell and WiFi subnetworks with unlicensed band access. The success (or outage) probabilities studied in all prior works are only for single-RAT cellular networks (typically see \cite{JGAFBRKG11}) so that their results cannot completely characterize the success probability of a multi-RAT network.  

Under the assumption that all APs use the carrier-sense-multiple access (CSMA) protocol to access all channels in the unlicensed band, our first contribution is to find the closed-form expression for the success probabilities of the two subnetworks and use them to characterize the coexisting success probability that is essentially the arithmetic average of all the success probabilities. To the best of our knowledge, the coexisting success probability is first proposed in this work. Our second contribution is to theoretically show that there exists an optimal ratio of the small cell density to the WiFi density that maximizes the coexisting success probability if the derived constraint on network parameters holds. Finally, the coexisting throughput that is the per-channel sum of all spectrum efficiencies of all subnetworks is proposed and numerical results show that it is significantly higher than the throughput of the unlicensed band only accessed by WiFi APs. 

\section{Network Model and Assumptions}
Consider a large-scale wireless network in which there coexist two  subnetworks of different RATs: one is a small cell (cellular) network and the other is a WiFi network. In the subnetwork of RAT ``$\mathsf{r}$'', all access points (APs) form a marked homogeneous Poisson point process (PPP) of density $\lambda_{\mathsf{r}}$ denoted by 
 \begin{align}
 \Phi_{\mathsf{r}} \defn\{(A_{\mathsf{r},i},H_{\mathsf{r},i},P_{\mathsf{r}}):A_{\mathsf{r},i}\in\mathbb{R}^2, P_{\mathsf{r}},H_{\mathsf{r},i}\in\mathbb{R}_+\},
 \end{align}
 where $\forall i\in\mathbb{N}_+$, $\mathsf{r}\in \mathcal{R}\defn\{\mathsf{s},\mathsf{w}\}$ in which ``$\mathsf{s}$'' denotes the small cell subnetwork and ``$\mathsf{w}$'' stands for the WiFi subnetwork, $P_{\mathsf{r}}$ is the transmit power of all  RAT-$\mathsf{r}$ APs. All $H_{\mathsf{r},i}$'s characterize the fading and/or shadowing channel power gains in the downlink between access points and their serving users and they are independent and identically distributed (i.i.d.) random variables with unit mean for all $i$ and $\mathsf{r}\in\mathcal{R}$. Note that the APs of the RAT-$\mathsf{r}$ subnetwork only provide service to the users in the same RAT-$\mathsf{r}$ subnetwork, e.g., small cell (WiFi) users only connect to small cell (WiFi) APs and cannot connect to  WiFi (small cell) APs.  
 
  
Assume there are $m$ channels in the unlicensed band available for all WiFi and small cell APs. All WiFi and small cell APs only try to access one of the $m$ channels by using the CSMA protocol\footnote{According to the recent 3GPP standard proposal, small cell APs are suggested to use the Listen-Before-Talk (LBT) protocol to access the channel in the unlicensed band. Since the idea of LBT is very similar to CSMA, here we assume small cell APs, like WiFi APs, also use CSMA to access the unlicensed band channels in order to simplify the channel access model and analysis.}. In this CSMA protocol, each AP has a circular sensing area in which it has to contend a channel with any other APs in this sensing area. If there are multiple channels available in the sensing area of an AP, the AP just equally likely accesses only one of them. Thus, each AP can retain a transmission opportunity if there is at least one channel available in its sensing area. Once APs grant a channel, they transmit immediately in the next transmitting time slot since we assume all APs always have data to transmit. The transmission probability of a WiFi or a small cell AP is shown in the following lemma.
    \begin{lemma}\label{Thm:TranProbWiFiAP}
  Assume every AP belonging to the RAT-$\mathsf{r}$ subnetwork is able to detect the sensing/transmitting behaviors of all other APs within its circular sensing area of radius $R_{\mathsf{r}}$.  The transmitting (retaining) probability of a AP is given by
   \begin{align}\label{Eqn:TransProbWiFiAP}
   \eta_{\mathsf{r}} =1-\left\{1-\frac{[1-\exp(-\pi R_{\mathsf{r}}^2\sum_{\mathsf{r}\in\mathcal{R}}\lambda_{\mathsf{r}}/m)]}{\pi R_{\mathsf{r}}^2 \sum_{\mathsf{r}\in\mathcal{R}}\lambda_{\mathsf{r}}/m}\right\}^m.
   \end{align} 
    \end{lemma}
  \begin{IEEEproof}
 Since the sensing area of an RAT-$\mathsf{r}$ AP is $\pi R^2_{\mathsf{r}}$, the average number of all different RAT APs in this area is $\pi R^2_{\mathsf{r}}\sum_{\mathsf{r}\in\mathcal{R}}\lambda_{\mathsf{r}}$. According to \cite{FBBBL10}, the probability that an RAT-$\mathsf{r}$ AP can access one of the $m$ channels is $p(m)=(1-e^{-\pi R^2_{\mathsf{r}}\sum_{\mathsf{r}\in\mathcal{R}}\lambda_{\mathsf{r}}/m})/(\pi R^2_{\mathsf{r}}\sum_{\mathsf{r}\in\mathcal{R}}\lambda_{\mathsf{r}}/m)$ if each AP tries to access one of the $m$ channels with equal probability. Hence, the probability that an RAT-$\mathsf{r}$ AP cannot access all $m$ channels is $[1-p(m)]^m$, which obviously indicates the probability that the AP can access at least one channel is $1-[1-p(m)]^m$ that is exactly equal to $ \eta_{\mathsf{r}}$ in \eqref{Eqn:TransProbWiFiAP}. 
  \end{IEEEproof} 
Due to the CSMA protocol, the resulting transmitting RAT-$\mathsf{r}$ APs form a  Mat\'{e}rn HCPP of density $\eta_{\mathsf{r}}\lambda_{\mathsf{r}}$ and theoretically the probability generating functional of a PPP cannot be applied to them because of the spacial correlation between them \cite{FBBBL10,DSWKJM96}. This exasperates the analysis of some performance metrics, such as success probability, average link rate, etc. Nonetheless, for the tractability of analysis \textit{we assume all transmitting RAT-$\mathsf{r}$ APs still form a homogeneous thinning PPP of density $\lambda_{\mathsf{r}}\eta_{\mathsf{r}} $}. Such an assumption is actually very accurate and valid when $\lambda_{\mathsf{r}}$ is not large. 

  
\section{Analysis of Coexisting Success Probability}  
Provided that every user belonging to the RAT-$\mathsf{r}$ subnetwork associates with an AP, its success probability based on its location at the origin for a particular channel is defined by
\begin{align}\label{Eqn:CovProb}
\rho_{\mathsf{r}}(\theta_{\mathsf{r}}) \defn \mathbb{P}\left[\frac{H_{\mathsf{r}}P_{\mathsf{r}}\|A_{\mathsf{r},0}\|^{-\alpha}}{\sum_{\mathsf{t}\in \mathcal{R}}I_{\mathsf{t}}}\geq\theta_{\mathsf{r}}\right],\,\, \mathsf{r}\in \mathcal{R},
\end{align}
where $\|X-Y\|$ denotes the Euclidean distance between two nodes $X$ and $Y$, $\alpha>2$ is the path loss exponent, $A_{\mathsf{r},0}$ is the associated AP of the user, $\theta_{\mathsf{r}}>0$ is the signal-to-interference ratio (SIR) threshold for successful decoding, and $I_{\mathsf{t}}$ is the interference from the RAT-$\mathsf{t}$ subnetwork given by
$$I_{\mathsf{t}}=\begin{cases}
\sum_{A_{\mathsf{r},i}\in\tilde{\Phi}_{\mathsf{r}}\setminus A_{\mathsf{r},0}}P_{\mathsf{r}} G_{\mathsf{r},i} \|A_{\mathsf{r},i}\|^{-\alpha}, &\text{if }\mathsf{t}=\mathsf{r}\\
\sum_{A_{\mathsf{r},i}\in\tilde{\Phi}_{\mathsf{r}}}P_{\mathsf{r}} G_{\mathsf{r},i} \|A_{\mathsf{r},i}\|^{-\alpha}, &\text{otherwise}
\end{cases},$$
where $\tilde{\Phi}_{\mathsf{r}}\subset\Phi_{\mathrm{r}}$ denotes the set of the transmitting APs using the particular channel in set $\Phi_{\mathrm{r}}$ and its density is $\eta_{\mathsf{r}}\lambda_{\mathsf{r}}/m$, $G_{\mathsf{r}}$'s represent the i.i.d. interference channel gains due to fading and/or shadowing and they have the same distribution as $H_i$'s.  

The success probability in \eqref{Eqn:CovProb} depends on how a user associates with its AP.  The following theorem renders an explicit result of $\rho_{\mathsf{r}}$ for nearest AP association\footnote{The reason that the nearest AP association scheme is considered in this paper is two-fold. First, it is more suitable for WiFi and small cell users whose channels are usually static and do not change too much along time. Second, it makes the derivation of the success probability much tractable.} and Rayleigh-fading in communication channels. 
\begin{theorem}\label{Thm:CovProb}
If all users belonging to the RAT-$\mathsf{r}$ subnetwork associate with their nearest AP and their communication channels undergo Rayleigh fading, the success probability in \eqref{Eqn:CovProb} is
\begin{align}
\rho_{\mathsf{r}}= \left\{1+\frac{\theta_{\mathsf{r}}^{\frac{2}{\alpha}}}{m}\left[\tau_{\alpha}\sum_{\mathsf{t}\in\mathcal{R}}\frac{\eta_{\mathsf{t}}\lambda_{\mathsf{t}}}{\eta_{\mathsf{r}}\lambda_{\mathsf{r}}}\left(\frac{P_{\mathsf{t}}}{P_{\mathsf{r}}}\right)^{\frac{2}{\alpha}}-\ell(\theta_{\mathsf{r}})\right]\right\}^{-1}, \label{Eqn:CovProbRATr}
\end{align}
where $\eta_{\mathsf{t}}$ is the transmitting probability of an RAT-$\mathsf{t}$ AP, $\ell(x)\defn \int_{0}^{x^{-\frac{2}{\alpha}}}(1-\mathcal{L}_G(t^{-\frac{\alpha}{2}}))\dif t$, $\mathcal{L}_Z(s)\defn\mathbb{E}[e^{-sG}]$ is the Laplace transform of random variable $Z$ and $\tau_{\alpha}\defn \Gamma\left(1-\frac{2}{\alpha}\right)\mathbb{E}[G^{\frac{2}{\alpha}}]$. Furthermore, if all interference channels also undergo Rayleigh, $\ell(x)=\int_{0}^{x^{-\frac{2}{\alpha}}} \frac{\dif t}{1+t^{\frac{\alpha}{2}}}$ and $\tau_{\alpha}=\Gamma(1-\frac{2}{\alpha})\Gamma(1+\frac{2}{\alpha})=2\pi\csc(2\pi/\alpha)/\alpha$.
\end{theorem}
\begin{IEEEproof}
Since the interferences from different RATs are independent and $H_{\mathsf{r}}$ is exponentially distributed with unit mean, the success probability in \eqref{Eqn:CovProb} can be rewritten as
\begin{align}
\rho_{\mathsf{r}}=\mathbb{E}_{D_{\mathsf{r}}}\left\{\prod_{\mathsf{t}\in\mathcal{R}}\mathbb{E}_{I_{\mathsf
		t}}\left[e^{-\theta_{\mathsf{r}} D^{\alpha}_{\mathsf{r}}I_{\mathsf{t}}/P_{\mathsf{r}}}\bigg|D_{\mathsf{r}}\right]\right\},\label{Eqn:CovProbInd}
\end{align}
where $D_{\mathsf{r}}$ denotes $\|A_{\mathsf{r},0}\|$. If $\mathsf{r}\neq\mathsf{t}$, $D_{\mathsf{r}}$ and $I_{\mathsf{t}}$ are independent and following the outage probability results in \cite{MHRKG09,CHLJGA12} gives
$$\mathbb{E}\left[e^{-\theta_{\mathsf{r}} D^{\alpha}_{\mathsf{r}}I_{\mathsf{t}}/P_{\mathsf{r}}}\bigg|D_{\mathsf{r}}\right]=\exp\left(-\pi D^2_{\mathsf{r}}\theta_{\mathsf{r}}^{\frac{2}{\alpha}}\tau_{\alpha}\eta_{\mathsf{t}}\frac{\lambda_{\mathsf{t}}}{m}\left(\frac{P_{\mathsf{t}}}{P_{\mathsf{r}}}\right)^{\frac{2}{\alpha}}\right).$$
However, if $\mathsf{r}=\mathsf{t}$, $D_{\mathsf{r}}$ and $I_{\mathsf{t}}$ are no longer independent and using the result in \cite{CHLLCW15} for a given $D_{\mathsf{r}}=x$ leads to
\begin{align*}
&\mathbb{E}\left[e^{-\theta_{\mathsf{r}} x^{\alpha}I_{\mathsf{r}}/P_{\mathsf{r}}}\right]=\mathbb{E}\left[\exp\left(-\theta_{\mathsf{r}}\sum_{A_{\mathsf{r},i}\in\Phi_{\mathsf{r}}\setminus A_{\mathsf{r},0}}\left(\frac{x^2G^{\frac{2}{\alpha}}_{\mathsf{r},i}}{\|A_{\mathsf{r},i}\|^2}\right)^{\frac{\alpha}{2}}\right)\right]\\
&= \mathbb{E}\left[\exp\left(-\theta_{\mathsf{r}}\sum_{A_{\mathsf{r},i}\in\Phi_{\mathsf{r}}\setminus A_{\mathsf{r},0}}\frac{x^{\alpha}G_{\mathsf{r},i}}{\left(x^2+\|A_{\mathsf{r},i}\|^2\right)^{\frac{\alpha}{2}}}\right)\right]\\
&=\exp\left(-\pi\eta_{\mathsf{r}}\lambda_{\mathsf{r}} x^2 \theta_{\mathsf{r}}^{\frac{2}{\alpha}}\left[\tau_{\alpha}-\ell(\theta_{\mathsf{r}})\right]/m \right).
\end{align*}
Then substituting the two above results into \eqref{Eqn:CovProbInd} yields 
\begin{align*}
\rho_{\mathsf{r}}=\mathbb{E}_{D_{\mathsf{r}}}\left[e^{-\pi \eta_{\mathsf{r}}\frac{\lambda_{\mathsf{r}}}{m} D_{\mathsf{r}}^2 \theta_{\mathsf{r}}^{\frac{2}{\alpha}}\left(\tau_{\alpha}\sum_{\mathsf{t}\in\mathcal{R}}\frac{\eta_{\mathsf{t}}\lambda_{\mathsf{t}}}{\eta_{\mathsf{r}}\lambda_{\mathsf{r}}}\left(\frac{P_{\mathsf{t}}}{P_{\mathsf{r}}}\right)^{\frac{2}{\alpha}}-\ell(\theta_{\mathsf{r}})\right)}\right].
\end{align*}
Since $A_{\mathsf{r},0}$ is the nearest AP of the user using RAT $\mathsf{r}$, the pdf of $D^2_{\mathsf{r}}$ is $f_{D^2_{\mathsf{r}}}(x)=\pi\eta_{\mathsf{r}}\lambda_{\mathsf{r}}e^{-\pi\eta_{\mathsf{r}}\lambda_{\mathsf{r}} x}$ and using it to calculate $\rho_{\mathsf{r}}$ in above results in \eqref{Eqn:CovProbRATr}. Finally, if all interference channels also undergo Rayleigh fading, $G$ is an exponential random variable with unit mean and variance and thus $\ell(x)$ and $\tau_{\alpha}$ can be further explicitly found.
\end{IEEEproof}

\begin{figure}[!t]
\centering
\includegraphics[width=3.5in, height=2.25in]{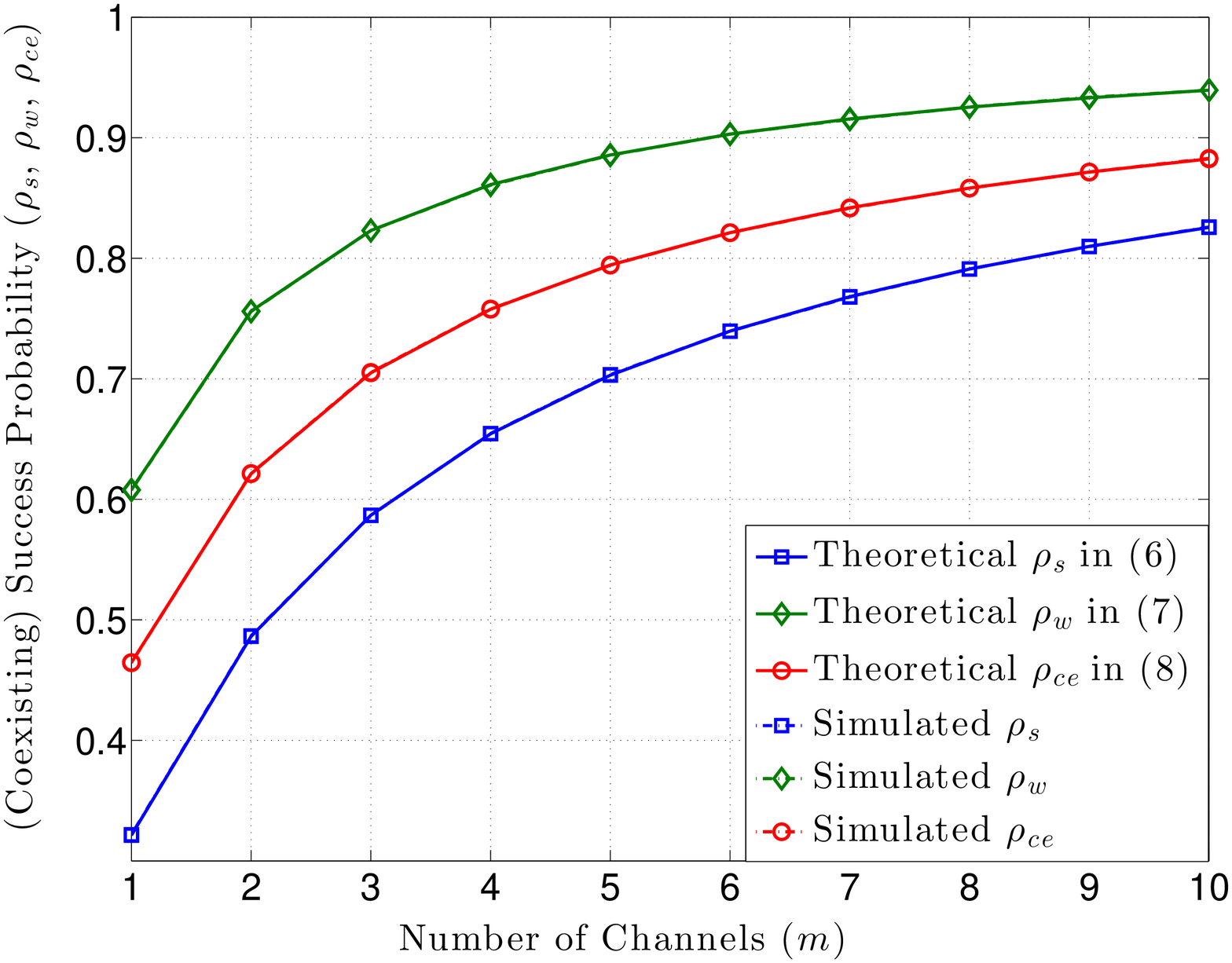}
\caption{Simulation results of the success and coexisting success probabilities for different numbers of channels. The RAT-$\mathsf
r$ APs are assumed as a homogeneous PPP of density $\eta_{\mathsf{r}}\lambda_{\mathsf{r}}$ in the network with Rayleigh fading and simulation parameters $\alpha=4$, $\lambda_{\mathsf{s}}=10 ^{-4}$ small cell APs/m$^2$, $\frac{\lambda_{\mathsf{w}}}{\lambda_{\mathsf{s}}}=3$, $P_{\mathsf{s}}=1$w, $P_{\mathsf{w}}=0.5$w, $R_{\mathsf{w}}=30$m, $R_{\mathsf{s}}=50$m, $\theta_{\mathsf{s}}=\theta_{\mathsf{w}}=0.5$.} 
\label{Fig:CoeCovProbMulCha}
\end{figure}

According to Theorem \ref{Thm:CovProb}, specifically the success probabilities for small cell and WiFi APs can be inferred and explicitly given as follows 
\begin{align}
&\hspace{-0.18in}\rho_{\mathsf{s}} = \left\{1+\frac{\theta_{\mathsf{s}}^{\frac{2}{\alpha}}}{m}\left[\tau_{\alpha} \left(1+\frac{\eta_{\mathsf{w}}\lambda_{\mathsf{w}}}{\eta_{\mathsf{s}}\lambda_{\mathsf{s}}}\left(\frac{ P_{\mathsf{w}}}{P_{\mathsf{s}}}\right)^{\frac{2}{\alpha}}\right)-\ell(\theta_{\mathsf{s}})\right]\right\}^{-1}, \label{Eqn:CovProbSmallCell}\\
&\hspace{-0.18in}\rho_{\mathsf{w}} = \left\{1+\frac{\theta_{\mathsf{w}}^{\frac{2}{\alpha}}}{m}\left[\tau_{\alpha} \left(1+\frac{\eta_{\mathsf{s}}\lambda_{\mathsf{s}}}{\eta_{\mathsf{w}}\lambda_{\mathsf{w}}}\left(\frac{P_{\mathsf{s}}}{P_{\mathsf{w}}}\right)^{\frac{2}{\alpha}}\right)-\ell(\theta_{\mathsf{w}})\right]\right\}^{-1}. \label{Eqn:CovProbWiFi}
\end{align}
To evaluate the coexisting success performance of the multi-RAT wireless network, the following \textbf{coexisting success probability} is proposed:
\begin{align}
\rho_{ce}\defn \frac{1}{|\mathcal{R}|}\sum_{\mathsf{r}\in\mathcal{R}}\rho_{\mathsf{r}},\label{Eqn:CoexistingCovProb}
\end{align}
where $|\mathcal{R}|$ represents the cardinality of set $\mathcal{R}$, i.e., the number of different RAT subnetworks. The coexisting success probability is essentially defined as the arithmetic average of all success probabilities in the network. 

As shown in \eqref{Eqn:TransProbWiFiAP}, the transmitting (retaining) probability $\eta_{\mathsf{r}}$ is a monotonically increasing function of $m$ and $\lim_{m\rightarrow\infty}\eta_{\mathsf{r}}=1$, which makes the coexisting success probability $\rho_{ce}$ also monotonically and concavely increases to one as $m$ goes to infinity, as shown in the following theorem.
\begin{theorem} \label{Thm:ConCaveCoeSuccProb}
For a given set of the deployment densities of all APs,  the coexisting success probability $\rho_{ce}$ in \eqref{Eqn:CoexistingCovProb} is a monotonically increasing and concave function of the channel number $m$ so that it increases up to one as $m$ goes to infinity.
\end{theorem}  	
\begin{IEEEproof}
Without loss of generality, the success probabilities $\rho_{\mathsf{s}}$ and $\rho_{\mathsf{w}}$ can be equivalently expressed as follows
\begin{align*}
\rho_{\mathsf{s}}=\frac{m}{m+a_{\mathsf{s}}\eta_{\mathsf{w}}/\eta_{\mathsf{s}}+b_{\mathsf{s}}}\,\,\text{ and }\,\, \rho_{\mathsf{w}}=\frac{m}{m+a_{\mathsf{w}}\eta_{\mathsf{s}}/\eta_{\mathsf{w}}+b_{\mathsf{w}}},
\end{align*}
where $a_{\mathsf{r}}$'s and $b_{\mathsf{r}}$'s are pertaining to $\lambda_{\mathsf{r}}$, $\theta_{\mathsf{r}}$, $P_{\mathsf{r}}$ and $\alpha$ for $\mathsf{r}\in\mathcal{R}$. Since we can show $\frac{\dif \rho_{\mathsf{r}}}{\dif m}>0$ for all $m$ and $\mathsf{r}\in\mathcal{R}$ and we know $\rho_{\mathsf{s}},\rho_{\mathsf{w}}\rightarrow 1$ as $m\rightarrow\infty$, $\rho_{\mathsf{s}}$ and $\rho_{\mathsf{w}}$ are a monotonically increasing and concave function of $m$. As a result, $\rho_{ce}$ is also a monotonically increasing and concave function of $m$ since it is a linear combination of $\rho_{\mathsf{w}}$ and $\rho_{\mathsf{s}}$. Obviously, $\rho_{ce}$ also increases to one as $m$ goes to infinity.
\end{IEEEproof}	

To illustrate the result of Theorem \ref{Thm:ConCaveCoeSuccProb},  the simulation and theoretical results of the success and coexisting success probabilities for different values of $m$ in a network with area 1 km$^2$ and other parameters are shown in Fig. \ref{Fig:CoeCovProbMulCha}. As expected, both of the success and coexisting success probabilities are a monotonically increasing and concave function of $m$. More importantly, the simulated results verify the validness of the theoretical results in \eqref{Eqn:CovProbSmallCell}, \eqref{Eqn:CovProbWiFi} and \eqref{Eqn:CoexistingCovProb}. In addition, the coexisting success probability can be optimized by selecting appropriate densities of APs.  The following theorem shows the fundamental constraints on the network parameters for achieving the optimality of the coexisting success probability with appropriate AP densities. 
\begin{theorem}\label{Thm:CoeExiProbOptDen}
For a given channel number $m$, the coexisting success probability $\rho_{ce}$ in \eqref{Eqn:CoexistingCovProb} can be maximized if the following inequality holds
\begin{align}
\min\left\{c_{\mathsf{w}}, c_{\mathsf{s}}\right\}>  \frac{(\theta_{\mathsf{s}}\theta_{\mathsf{w}})^{\frac{1}{\alpha}}\tau_{\alpha}}{m},\label{Eqn:OptDenCons}
\end{align}
where $c_{\mathsf{r}}\defn 1+\theta_{\mathsf{r}}^{\frac{2}{\alpha}}\left[\tau_{\alpha}-\ell(\theta_{\mathsf{r}})\right]/m$,  $\forall\mathsf{r}\in\mathcal{R}$, and the optimal density ratio of $\eta_{\mathsf{s}}\lambda_{\mathsf{s}}$ to $\eta_{\mathsf{w}}\lambda_{\mathsf{w}}$ is equal to  
\begin{align}
\left(\frac{\eta_{\mathsf{s}}\lambda_{\mathsf{s}}}{\eta_{\mathsf{w}}\lambda_{\mathsf{w}}}\right)^*=\left(\frac{\sqrt{\theta_{\mathsf{s}}}P_{\mathsf{w}}}{\sqrt{\theta_{\mathsf{w}}}P_{\mathsf{s}}}\right)^{\frac{2}{\alpha}}\left(\frac{mc_{\mathsf{w}}-\tau_{\alpha}(\theta_{\mathsf{s}}\theta_{\mathsf{w}})^{\frac{1}{\alpha}}}{mc_{\mathsf{s}}-\tau_{\alpha}(\theta_{\mathsf{s}}\theta_{\mathsf{w}})^{\frac{1}{\alpha}}}\right).\label{Eqn:OptDen1}
\end{align}
For the special case of $\theta_{\mathsf{r}}=\theta$ for all $\mathsf{r}\in\mathcal{R}$, the constraint \eqref{Eqn:OptDenCons} always holds for all $\theta>0$
and \eqref{Eqn:OptDen1} reduces to
\begin{align}
\left(\frac{\eta_{\mathsf{s}}\lambda_{\mathsf{s}}}{\eta_{\mathsf{w}}\lambda_{\mathsf{w}}}\right)^*=\left(\frac{P_{\mathsf{w}}}{P_{\mathsf{s}}}\right)^{\frac{2}{\alpha}}.\label{Eqn:OptDen2}
\end{align}
\end{theorem}
\begin{IEEEproof}
Using the definition of $c_{\mathsf{r}}$, the coexisting success probability $\rho_{ce}$ in \eqref{Eqn:CoexistingCovProb} can be concisely written as
$$\rho_{ce}(y)=\frac{y/2}{c_{\mathsf{s}} y+d_{\mathsf{s}}}+\frac{1/2}{c_{\mathsf{w}}+d_{\mathsf{w}}y}=\frac{1}{2c_{\mathsf{s}}}-\frac{d_{\mathsf{s}}/2c^2_{\mathsf{s}}}{y+d_{\mathsf{s}}/c_{\mathsf{s}}}+\frac{1/2d_{\mathsf{w}}}{y+c_{\mathsf{w}}/d_{\mathsf{w}}},$$
where $y\defn (\frac{\eta_{\mathsf{s}}\lambda_{\mathsf{s}}}{\lambda_{\mathsf{w}}\lambda_{\mathsf{w}}})(\frac{P_{\mathsf{s}}}{P_{\mathsf{w}}})^{\frac{2}{\alpha}}$, $d_{\mathsf{s}}\defn\frac{\tau_{\alpha}\theta_{\mathsf{s}}^{\frac{2}{\alpha}}}{m}$  and $d_{\mathsf{w}}\defn\frac{\tau_{\alpha}\theta_{\mathsf{w}}^{\frac{2}{\alpha}}}{m}$. Its $n$th derivative w.r.t. $y$ is found as
\begin{align*}
\rho^{(n)}_{ce}(y) =\frac{(-1)^{n+1}d_{\mathsf{s}}n! }{2c^2_{\mathsf{s}}(y+d_{\mathsf{s}}/c_{\mathsf{s}})^{n+1}}+\frac{(-1)^n n!}{2d_{\mathsf{w}}(y+c_{\mathsf{w}}/d_{\mathsf{w}})^{n+1}}.
\end{align*}
Hence,  the set $\{y\in\mathbb{R}_{++}: \rho^{(1)}_{ce}(y)=0, \rho^{(2)}_{ce}(y)<0 \}$ is not empty provided that $c^2_{\mathsf{s}}>d_{\mathsf{w}}d_{\mathsf{s}}$, $c^2_{\mathsf{w}}>d_{\mathsf{w}}d_{\mathsf{s}}$ and $\sqrt{c}_{\mathsf{s}}c^{\frac{3}{2}}_{\mathsf{w}}>d_{\mathsf{w}}d_{\mathsf{s}}$ hold (i.e. the constraint in \eqref{Eqn:OptDenCons} holds). The optimal value of $y$ can be solved from $\rho^{(1)}_{ce}(y)=0$ under condition \eqref{Eqn:OptDenCons}, which generates the result in \eqref{Eqn:OptDen1}. Also, in the case of $\theta_{\mathsf{r}}=\theta$ for all $\mathsf{r}\in\mathcal{R}$, the inequality in \eqref{Eqn:OptDenCons} reduces to $m>\ell(\theta)$, which is always valid since $m\geq 1$ and $\ell(\theta)<1$ for all $\theta>0$, and the result in \eqref{Eqn:OptDen2} can be obtained by substituting $c_{\mathsf{r}}$'s and $d_{\mathsf{r}}$'s with $\theta_{\mathsf{r}}=\theta$ into \eqref{Eqn:OptDen1}. This completes the proof. 
\end{IEEEproof}	

\begin{figure}[!t]
	\centering
	\includegraphics[width=3.5in, height=2.25in]{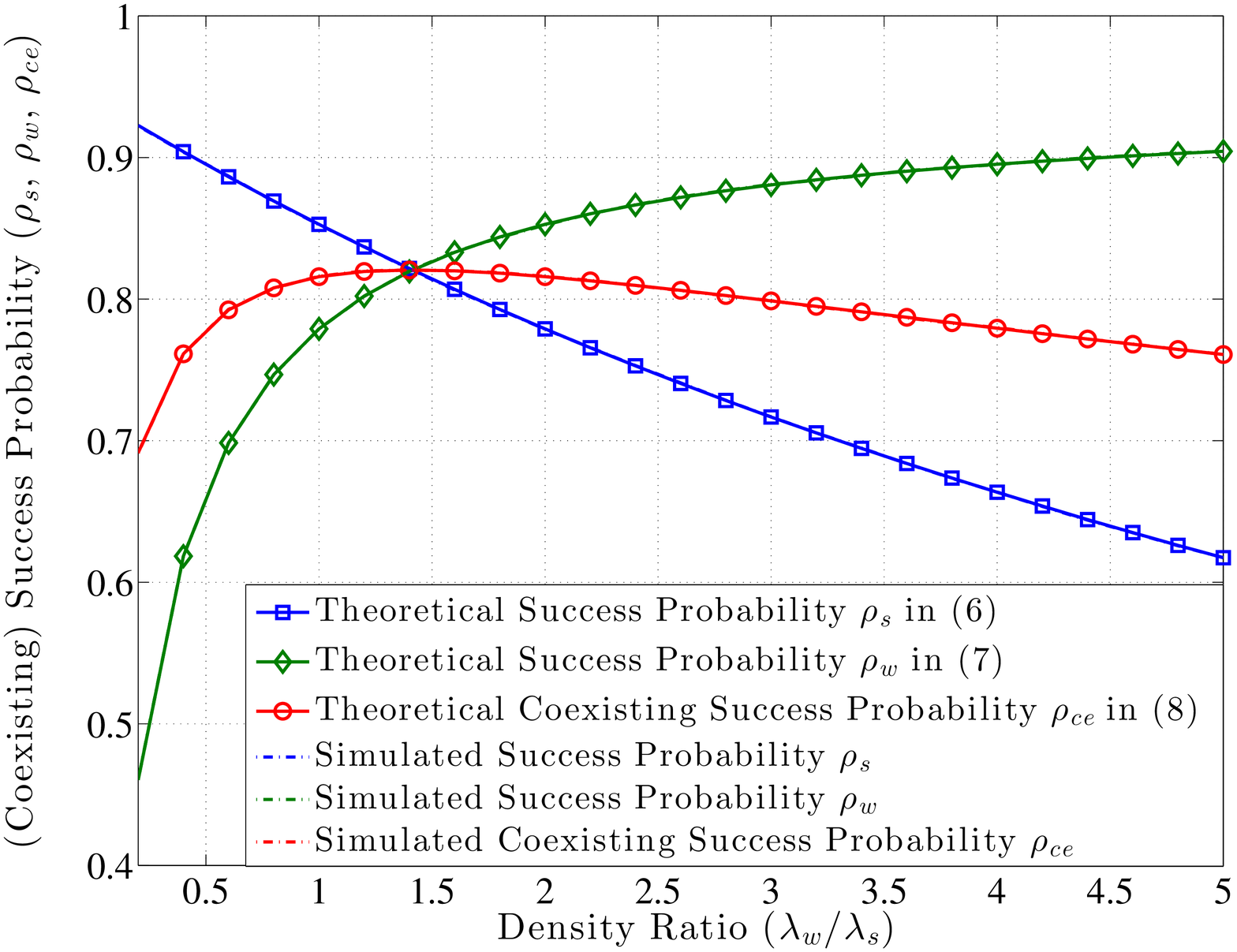}
	\caption{Simulation results of the success and coexisting success probabilities for different ratios of $\lambda_{\mathsf{w}}$ to $\lambda_{\mathsf{s}}$. The RAT-$\mathsf
	r$ APs are assumed as a homogeneous PPP of density $\eta_{\mathsf{r}}\lambda_{\mathsf{r}}$ in the network with Rayleigh fading and the following simulation parameters: network area 1 km$^2$, $\alpha=4$, $\lambda_{\mathsf{s}}=10 ^{-4}$ small cell APs/m$^2$, $m=5$, $P_{\mathsf{s}}=1$w, $P_{\mathsf{w}}=0.5$w, $R_{\mathsf{w}}=30$m, $R_{\mathsf{s}}=50$m, $\theta_{\mathsf{s}}=\theta_{\mathsf{w}}=0.5$.} 
	\label{Fig:CoeCovProb}
\end{figure}

The inequality constraint in \eqref{Eqn:OptDenCons} provides us some insights into the  design of the network parameters including the number of channels, deployment densities and SIR thresholds such that there exists a set of deployment densities to maximize the coexisting success probability. The simulation and theoretical results of the coexisting success probability for different density ratios and some given network parameters are illustrated in Fig. \ref{Fig:CoeCovProb}. Since $\theta_{\mathsf{s}}=\theta_{\mathsf{w}}=\theta=0.5$, the constraint \eqref{Eqn:OptDenCons} is always true for all $\theta>0$ and the optimal ratio of $\lambda_{\mathsf{w}}$ to $\lambda_{\mathsf{s}}$ for the simulation network parameters can be numerically found as $\left(\frac{\lambda_{\mathsf{w}}}{\mathsf{s}}\right)^*=1.4$ by using \eqref{Eqn:OptDen2} with $\lambda_{\mathsf{s}}=10^{-4}$, which coincides with the simulation result in Fig.  \ref{Fig:CoeCovProb}.  Hence, there indeed exists (at least) one pair of densities $\lambda_{\mathsf{s}}$ and $\lambda_{\mathsf{w}}$ that maximizes $\rho_{ce}$.


\section{Coexisting Throughput and Its Optimality}
According to the definition of the coexisting success probability,  in a multi-RAT wireless network with $m$ channels the following \textbf{coexisting throughput} (bps/Hz/channel) is proposed and defined as
\begin{align}
\mathsf{C}_{ce} &\defn\frac{1}{m}\sum_{\mathsf{r}\in\mathcal{R}}\mathbb{E}\left[\log_2\left(1+\frac{H_{\mathsf{r}}P_{\mathsf{r}}\|A_{\mathsf{r},0}\|^{-\alpha}}{\sum_{\mathsf{t}\in \mathcal{R}}I_{\mathsf{t}}}\right)\right]\nonumber\\
&=\frac{1}{m}\sum_{\mathsf{r}\in\mathcal{R}}\int_0^{\infty} \rho_{\mathsf{r}}\left(2^{x}-1\right) \dif x,
\end{align}
where the last equality is due to $\mathbb{E}[\log_2(1+X)]=\int_0^{\infty} \mathbb{P}[\log_2(1+X)\geq x]\dif x=\int_0^{\infty} \mathbb{P}[X\geq 2^x-1]\dif x$ and using $\rho_{\mathsf{r}}$ in \eqref{Eqn:CovProb}.  Essentially, $\mathsf{C}_{ce}$ is the per-channel sum of the spectrum efficiencies of all different RAT subnetworks. 

\begin{figure}[!t]
	\centering
	\includegraphics[width=3.5in, height=2.25in]{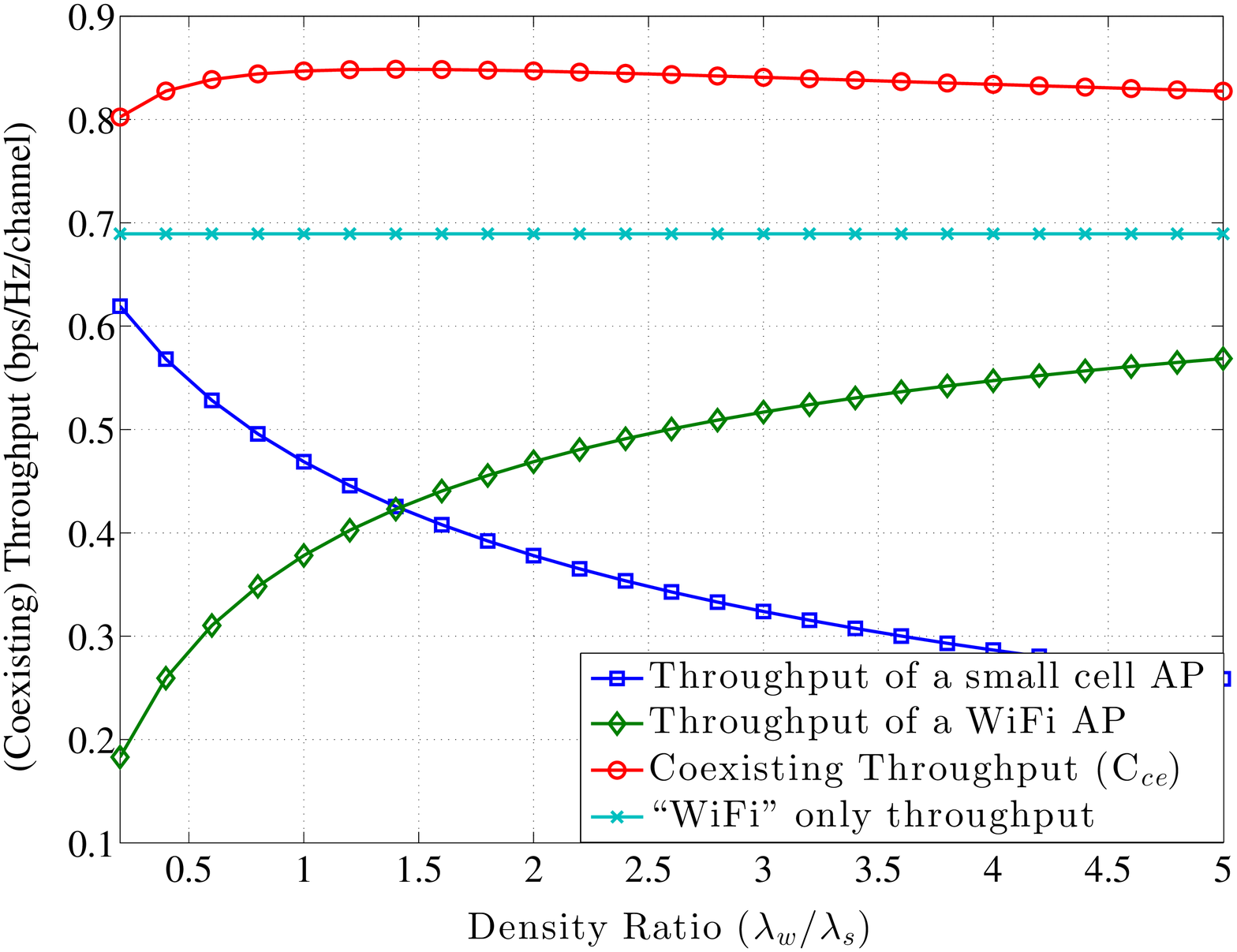}
	\caption{Simulation results of the coexisting throughput and different RAT throughputs.  The RAT-$\mathsf
		r$ APs are assumed as a homogeneous PPP of density $\eta_{\mathsf{r}}\lambda_{\mathsf{r}}$ in the network with Rayleigh fading and the following simulation parameters: network area 1 km$^2$, $\alpha=4$, $\lambda_{\mathsf{s}}=10 ^{-4}$ small cell APs/m$^2$, $m=5$, $P_{\mathsf{s}}=1$w, $P_{\mathsf{w}}=0.5$w, $R_{\mathsf{w}}=30$m, $R_{\mathsf{s}}=50$m, $\theta_{\mathsf{s}}=\theta_{\mathsf{w}}=0.5$.} 
	\label{Fig:MeanSumRate}
\end{figure}

According to Theorems \ref{Thm:ConCaveCoeSuccProb} and \ref{Thm:CoeExiProbOptDen}, we can infer that the coexisting throughput $\mathsf{C}_{ce}$ can be jointly optimized over channel number $m$ and density ratio $\frac{\lambda_{\mathsf{s}}}{\lambda_{\mathsf{w}}}$ as well.  However, its optimal results are analytically intractable. Nevertheless, Fig. \ref{Fig:MeanSumRate} shows the simulation results of the small cell, WiFi and coexisting throughputs for a fixed $m$ and it verifies the optimality of $\mathsf{C}_{ce}$ over density ratio $\frac{\lambda_{\mathsf{s}}}{\lambda_{\mathsf{w}}}$. As shown in the figure, the ``WiFi'' only throughput, i.e., the throughput of the unlicensed band only accessed by WiFi APs, is a constant around 0.689 (bps/Hz/channel), whereas  coexisting throughput $\mathsf{C}_{ce}$ is between 0.8 and 0.85. Thus, the throughput gain due to coexistence in the unlicensed band for the network is around 22\%. Most importantly, there indeed exists an optimal density ratio $(\frac{\lambda_{\mathsf{w}}}{\lambda_{\mathsf{s}}})^* \approx 1.45$ that maximizes $\mathsf{C}_{ce}$ and achieves about 30\% throughput gain, which is a fairly significant improvement in throughput.

\section{Conclusion}
The coexisting success probability is first investigated for a wireless network specifically consisting of two WiFi and small cell subnetworks. For given deployment densities, it is a monotonically increasing and concave function of the number of channels, whereas the optimal ratio of the AP densities maximizing it is shown to exist and found for a given number of channels if the AP densities satisfy the derived constraint on the network parameters.  The coexisting throughput characterized by the sum of all the spectrum efficiencies of all subnetworks is proposed and numerical results show that it is significantly larger than the throughput of the unlicensed band accessed only by WiFi APs and can be jointly maximized over the AP densities as well as the number of channels.

\bibliographystyle{ieeetran}
\bibliography{IEEEabrv,Ref_ThrPutHetUnliband}

\begin{thebibliography}{1}
\providecommand{\url}[1]{#1}
\csname url@samestyle\endcsname
\providecommand{\newblock}{\relax}
\providecommand{\bibinfo}[2]{#2}
\providecommand{\BIBentrySTDinterwordspacing}{\spaceskip=0pt\relax}
\providecommand{\BIBentryALTinterwordstretchfactor}{4}
\providecommand{\BIBentryALTinterwordspacing}{\spaceskip=\fontdimen2\font plus
\BIBentryALTinterwordstretchfactor\fontdimen3\font minus
  \fontdimen4\font\relax}
\providecommand{\BIBforeignlanguage}[2]{{%
\expandafter\ifx\csname l@#1\endcsname\relax
\typeout{** WARNING: IEEEtran.bst: No hyphenation pattern has been}%
\typeout{** loaded for the language `#1'. Using the pattern for}%
\typeout{** the default language instead.}%
\else
\language=\csname l@#1\endcsname
\fi
#2}}
\providecommand{\BIBdecl}{\relax}
\BIBdecl

\bibitem{HZXCWGSW15}
H.~Zhang, X.~Chu, W.~Guo, and S.~Wang, ``Coexistence of {W}i-{F}i and
  heterogeneous small cell networks sharing unlicensed spectrum,'' \emph{{IEEE}
  Commun. Mag.}, vol.~53, no.~3, pp. 158--164, Mar. 2015.

\bibitem{JGAFBRKG11}
J.~G. Andrews, F.~Baccelli, and R.~K. Ganti, ``A tractable approach to coverage
  and rate in cellular networks,'' \emph{{IEEE} Trans. Commun.}, vol.~59,
  no.~11, pp. 3122--3134, 2011.

\bibitem{FBBBL10}
F.~Baccelli and B.~B{\l}aszczyszyn, ``Stochastic geometry and wireless
  networks: Volume {II A}pplications,'' \emph{Foundations and Trends in
  Networking}, vol.~3, no. 3-4, pp. 249--449, 2010.

\bibitem{DSWKJM96}
D.~Stoyan, W.~Kendall, and J.~Mecke, \emph{Stochastic Geometry and Its
  Applications}, 2nd~ed.\hskip 1em plus 0.5em minus 0.4em\relax New York: John
  Wiley and Sons, Inc., 1996.

\bibitem{MHRKG09}
M.~Haenggi and R.~K. Ganti, ``Interference in large wireless networks,''
  \emph{Foundations and Trends in Networking}, vol.~3, no.~2, pp. 127--248,
  2009.

\bibitem{CHLJGA12}
C.-H. Liu and J.~G. Andrews, ``Ergodic transmission capacity of wireless ad hoc
  networks with interference management,'' \emph{{IEEE} Trans. Wireless
  Commun.}, vol.~11, no.~6, pp. 2136--2147, Jun. 2012.

\bibitem{CHLLCW15}
C.-H. Liu and L.-C. Wang, ``Optimal cell load and throughput in green small
  cell networks with generalized cell association,'' \emph{submitted for
  publication (available at http://arxiv.org/abs/1503.08661)}, Mar. 2015.

\end{thebibliography}

\end{document}